\documentclass[twocolumn,prc,showpacs,preprintnumbers,amsmath,amssymb]{revtex4}

\usepackage{graphicx}

\begin{document}

\title{Relativistic Effects in 3-Nucleon Forces for Nuclear Matter and Finite Nuclei}
\author{A.H. Lippok and H. M\"uther}
\affiliation{Institut f\"ur Theoretische Physik, \\
Universit\"at T\"ubingen, D-72076 T\"ubingen, Germany}

\begin{abstract}
In order to simulate the relativistic effects of the Dirac Brueckner Hartree Fock approach for finite nuclei the part of the Urbana 3 nucleon (3N) force is considered, which represents the enhancement of the small components of the Dirac spinors for the nucleons in the nuclear medium. This 3N force is included in a Brueckner Hartree Fock calculation with rearrangement terms using a realistic model for the NN interaction. The strength of the 3N force is adjusted to reproduce the empirical saturation point of nuclear matter and then used in corresponding studies of the closed shell nuclei $^{16}$O and $^{40}$Ca. Special attention is paid to a consistent treatment of the spectrum of particle states in the NN propagator of the Bethe-Goldstone equation.
\end{abstract}


\maketitle

\section{Introduction}

One of the main challenges of theoretical nuclear physics is the attempt to
derive the bulk properties of nuclear systems, which includes the saturation 
properties of infinite nuclear matter but also the binding energies and
size of finite nuclei, from a realistic model of the nucleon-nucleon (NN) interaction.
In this context models of the NN interactions are called realistic if they 
have been determined to fit the experimental data for 2 nucleons, i.e. the
NN scattering phase shifts and the data of the deuteron, with high accuracy.
These so-called realistic NN interactions are in contrast to phenomenological
or effective NN forces like the Skyrme interactions \cite{skyrme1,skyrme2} or relativistic mean field
models\cite{walecka}, which are fitted to describe the bulk properties of nuclear systems using
Hartree-Fock or mean-field approximations.

Examples for such realistic NN interactions are the One-Boson-Exchange (OBE) Models
of the Bonn (respectively Idaho)\cite{mach0} group or the local interaction models of the
Argonne or Urbana groups\cite{V18}. Rather sophisticated versions of these models
have been developed like the CDBONN potential\cite{cdbonn} or the Argonne V18 potential\cite{V18}.

Although these potentials are ``soft'' as compared to the hard-core potentials developed in the 
middle of the last century, they contain strong tensor- and short-range components, which
make it inevitable to employ a calculation tool which treats correlations beyond
the mean-field or Hartree-Fock approximation\cite{polls}. A typical example of such
a many-body approach is the Brueckner-Hartree-Fock (BHF) approximation, which is based 
on the solution of two-nucleon scattering equation in the nuclear medium, leading to
an energy- and density dependent effective interaction, the so-called G-matrix. 

Attempts have been made to compare this G-matrix with the phenomenological models of 
a nuclear force as the Skyrme interaction mentioned above and identify the density
dependence of the Skyrme interaction with the medium dependence of the G-matrix\cite{negele,eigenearb}.

The variation principle of Hartree-Fock calculations with density-dependent forces
leads to rearrangement terms, which in the case of the Skyrme interaction are very
important to obtain good agreement with the empirical data for nuclear matter and
finite nuclei. Therefore also the BHF approximation has been extended to include such
rearrangement terms. Accounting for the energy dependence of G this leads to the 
Renormalized BHF (RBHF) approach\cite{david} whereas the Density dependent HF (DHF) approximation
also accounts for the Pauli-rearrangement terms\cite{eigenearb,zuo}. The inclusion of the
rearrangement terms is not only justified to simulate the features contained in the effective
theory, but, in contrast to BHF, the DHF approach fulfills the Hughenholtz van Hove theorem
due to the inclusion of the rearrangement terms.\cite{Baymk} 
 
The BHF approach using realistic NN interaction leads to a saturation point for 
nuclear matter. The saturation points calculated with various models for the realistic NN interaction form the so-called Coester band\cite{polls,coester},
which misses the empirical data in a significant way.  Therefore many attempts
have been made to explore various effects, which make the NN interaction in the nuclear medium different from the NN interaction in the vacuum.
An example of such studies is the inclusion of sub-nucleonic degrees of freedom e.g. in
terms of intermediate excitations of the interacting nucleons to the $\Delta(3,3)$ resonance.
Such mutual polarization effects should occur already in the interaction of 2 nucleons
in the vacuum. In fact such processes with intermediate isobar excitations provide
a substantial contribution to the medium range attraction of the NN interaction, which
in the OBE model is described in terms of the exchange of a light scalar $\sigma$ meson\cite{holinde,wiringa}.
In the nuclear medium these attractive terms are quenched due to dispersive and
Pauli effects, features which could be described in terms of a density-dependent
NN force or a 3-N interaction\cite{delta1,delta2}. Such isobar effects turned out
to be non-negligible, however, did not cure the problem of the Coester band, i.e. did
not shift the saturation point calculated for nuclear matter to the empirical data.

Another form of medium-dependence of the NN interaction has been supplied by the
Dirac-Brueckner-Hartree-Fock (DBHF) approximation\cite{dirac1,dirac2}. Here one considers 
the Dirac structure of the self-energy of the nucleon and accounts for the effect
that the attractive scalar component of this self-energy yields Dirac spinors
for the nucleons in the nuclear medium with an enhanced small component as compared
to the corresponding spinor for a nucleon in the vacuum. Again this change of the
Dirac spinors in the medium and the corresponding change for the matrix elements of 
meson exchange can either be described in terms of a medium-dependent NN force or by
means of a 3N force parameterized in form of a Z-diagram with 2 $\sigma$ exchange terms\cite{brown87,Bouyssy87}.

The DBHF approximation has been  successful in the sense that it reproduced the
empirical saturation point of nuclear matter without adjustment of any additional parameter\cite{dirac1,anastasio83,horow87,terhaar:1987,dejong:1998,
gross:1999,alonso:2003,eric}. Since however the consistent treatment of
correlation  and relativistic effects for finite systems is a rather involved
problem, the full Dirac Brueckner equations have not yet been solved for finite nuclei.
In fact, Van Giai et al.~\cite{giai2010} addressed this problem as one
of the main open problems in Nuclear Physics.   Different approximation schemes
have been developed, which treat either the relativistic effects or the
correlation effects in an approximate way.

The most popular approximation scheme is to analyze the DBHF calculations of infinite matter in terms of
an effective field theory with meson-nucleon coupling constants depending on the nucleon density and 
perform relativistic mean field or  relativistic Hartree-Fock calculations for finite nuclei using the semi-phenomenological
density functionals, which were adjusted to reproduce DBHF results for infinite matter\cite{releric,fuchs95}. For a recent discussion of 
such relativistic density functionals see \cite{eric}  and references listed there. Such studies consider the relativistic features of DBHF explicitly, but ignore more or less the effects of correlations beyond the mean field approach.  It has been pointed out that the rearrangement terms originating from the density dependence of the meson-nucleon coupling constants
are very important to improve the description of finite nuclei\cite{fuchs95}. 

In the present investigation we are going to discuss a different approximation scheme for the DBHF approach in finite nuclei. The focus is to consider the correlation effects in terms of a direct evaluation for finite nuclei and treat the Dirac effects in an approximate way. However, in contrast to the method developed in \cite{brock88} the Dirac effects are not described in terms of a local density approximation but are simulated using a three-nucleon interaction.

Various attempts have been made to parameterize the 3N forces discussed so far in terms of
a simple local 3N force. As an example we mention the Urbana force\cite{urbana3,baldof}, which is composed
of two terms
\begin{equation}
V_{ijk} = A \,V_{ijk}^{2\pi} + U\,V_{ijk}^R\,.\label{eq:3-v}
\end{equation}
The first part is from $2\pi$ exchange with an intermediate $\Delta$ excitation and may 
be considered to simulate the medium-dependent isobar effects discussed above. The second
term is typically defined in terms of $2\sigma$ exchange and can be interpreted to simulate
the effects of the $Z$-diagram discussed above. This means the second term is thought to
represent the relativistic effects of the DBHF approach\cite{grange}. Typically this 3N force
is reduced to a density-dependent NN interaction, which is then added to the bare NN interaction
(see e.g. \cite{soma} and references therein) and the parameters $A$ and $U$ in eq.(\ref{eq:3-v})
can be adjusted to reproduce the empirical saturation point for symmetric nuclear matter.

This scheme has been criticized bey Hebeler and Schwenk\cite{hebschw} and later by Carbonne et al.\cite{carbone}
They argue that an expression for the total energy with kinetic energy $t_i$, 2N interaction $V_{ij}$ and
3N potential $V_{ijk}$
\begin{equation}
E = \sum_i t_i\rho_i + \frac{1}{2}\sum_{i,j} V_{ij}\rho_i \rho_j + \frac{1}{6} \sum_{i,j,k} V_{ikj}\rho_i \rho_k \rho_j\label{eq:3n1}
\end{equation}
leads to the single-particle energy
\begin{equation}
\varepsilon_i = t_i + \sum_j V_{ij}\rho_j + \frac{1}{2} \sum_{j,k} V_{ikj}\rho_k \rho_j\,,\label{eq:3n2}
\end{equation}
which is different from the result, which is obtained, when the 3N force is added to the 2N interaction by
\begin{equation}
\frac{1}{2} V_{ij}^{eff}(\rho )  = \frac{1}{2} V_{ij} + \frac{1}{6}\sum_k V_{ikj}\rho_k\label{eq:v3dens}
\end{equation}
This is of course true and at first sight this would imply that the medium effects discussed above would
lead to different results when they are treated in terms of a 3-body force or considered as a density-dependent
NN interaction. We note, however, that both approaches lead to the same result, if  the single-particle
energies are defined according to the Landau definition of the quasiparticle energy, i.e.
\begin{equation}
\varepsilon_i = \frac{\partial}{\partial \rho_i} E(\rho)\,,\label{eq:landaud}
\end{equation}
which means that rearrangement terms due to the density dependence of $V^{eff}$ are taken into account. With this
inclusion the result is the independent on the treatment as a 3N term or a density-dependent 2N contribution.

In this investigation we will discuss Brueckner-Hartree-Fock kind of calculations for nuclear matter and finite nuclei based on a realistic NN interaction with inclusion of a 3-nucleon force. Special attention will be paid to the rearrangement terms originating from the density dependence of the $G$-matrix and the treatment of the 3N force in terms of a density-dependent 2N interaction. We will show that an adjustment of the constant $U$ defining the relativistic 3-body force in (\ref{eq:3-v}) is sufficient to obtain the empirical saturation point for symmetric nuclear matter. The same 3N force leads to a fair description also for the bulk properties of finite nuclei.

After this introduction we will discuss the Brueckner Hartree Fock approach with 3N forces and inclusion of rearrangement terms in section 2 of this paper. The results for infinite matter and finite nuclei are discussed in section 3. Special attention will be paid to the treatment of the 3N force in finite nuclei and the description of the particle state spectrum in the Bethe-Goldstone equation. The main results and conclusions are summarized in section 4.

\section{Brueckner Hartree Fock and Rearrangement Terms}

\subsection{Nuclear Matter and Finite Nuclei}

The Brueckner Hartree Fock (BHF) approach can be defined in terms of three central equations. The first one of these equations is the Bethe-Goldstone equation
\begin{equation}
G(\omega) = V + V \frac{\hat Q}{\omega - \hat H_0} G\label{eq:betheg}
\end{equation} 
defining the so-called $G$ matrix in terms of the free-space NN interaction $V$. Replacing the Pauli Operator $\hat Q$, which forbids the scattering of the interacting nucleons into states, which are below the Fermi energy and therefore occupied by other nucleons by the unit operator and the energy denominator $\omega - \hat H_0$ by the difference of kinetic energies of free nucleons, the $G$ matrix turns over into the Lippman Schwinger equation defining the scattering matrix $T$ for two nucleons in the vacuum. Therefore the Bethe-Goldstone equation can be interpreted as the solution of the problem of two nucleons interacting in the nuclear medium and the $G$ matrix can be understood as the effective interaction of two nucleons, which accounts for correlation between the interacting nucleons. The single-particle energies
are then defined within the standard BHF using the Hartree-Fock expression in terms of the $G$-matrix interaction
\begin{equation}
\varepsilon_i^{BHF} = \langle i \vert \hat t \vert i \rangle + \sum_{j} \langle ij \vert G(\omega=\varepsilon_i + \varepsilon_j)\vert ij \rangle\rho_j\,.\label{eq:bhfeps}
\end{equation}
The single-particle density $\rho_j$ is diagonal in the basis of Hartree-Fock states 
$$
\langle k \vert \hat \rho \vert j \rangle = \rho_j \delta_{jk} \,,
$$
and the diagonal elements $\rho_j$  take the values 1 for occupied hole states with energies $\varepsilon_j$ below the Fermi energy and 0 for the particle states above the Fermi level. In the case of finite nuclei these Hartree Fock states have
to be determined as the eigenstates of the BHF single-particle Hamiltonian, which corresponds to the single-particle energies
defined in eq. (\ref{eq:bhfeps}). In the case of infinite nuclear matter these single-particle states are plane waves due to the symmetry of the system under translational transformation. Note, however that also in the case of infinite matter a self-consistent solution of the Bethe-Goldstone equation (\ref{eq:betheg}) and the evaluation of the single-particle energies according eq. (\ref{eq:bhfeps}) is required to determine the starting energy $\omega$ according to the Bethe-Brandow-Petchek theorem.

After solving eqs.(\ref{eq:betheg}) and (\ref{eq:bhfeps}) in a self-consistent way one can evaluate the total energy as
\begin{equation}
E = \sum_i  \langle i \vert\hat t \vert i \rangle \rho_i + \frac12 \sum_{ij} \langle ij \vert G(\omega=\varepsilon_i + \varepsilon_j)\vert ij \rangle\rho_j\rho_i\,,\label{eq:tote}
\end{equation}
which is the third of the 3 equations to define the BHF approximation. Note that using the BHF definition of the single-particle energies (\ref{eq:bhfeps}) the corresponding expression for the total energy leads to Koltuns sum rule\cite{koltun}
\begin{equation}
E^{BHF} = \sum_i \frac12\left(  \langle i \vert \hat t \vert i \rangle + \varepsilon_i^{BHF}\right) \rho_i
\label{eq:koltun}
\end{equation}

The definition of the single-particle energies for the intermediate particle states, i.e. the definition of the operator $H_0$ in
the propagator of the Bethe-Goldstone eq.(\ref{eq:betheg}) has been discussed for many years. It has been shown by Song et al.\cite{song} that the contribution from three-body 
correlations is minimized in nuclear matter with the so-called continuous prescription \cite{jeuken76}, which means that the single-particle energies for the states above the Fermi level are calculated in the same way as those for the hole states below. As will be discussed below, we will try to adopt this prescription in our calculations. Since it is very elaborate to evaluate the single-particle energies for all states in finite nuclei, we will approximate the Hamiltonian for the two particle states in this case by
\begin{equation}
\hat H_0 = \hat Q\left( \hat t_1 + \hat t_2 \right)\hat Q - 2C\,,\label{eq:h20}
\end{equation}
which is the operator of the kinetic energy of the interacting particles restricted to the states above the Fermi energy.
A constant $C$ is introduced to make the spectrum ``continuous''  across the Fermi energy. Appropriate values will be discussed below.

Another rather technical obstacle for the solution of the Bethe-Goldstone eq.(\ref{eq:betheg}) is the definition of the Pauli operator $\hat Q$. This Pauli operator is easily defined for nuclear matter using the rest-frame of the nuclear matter system by
$$
\hat Q \vert \vec k_1,\,\vec k_2 \rangle = \begin{cases} \vert\vec k_1,\,\vec k_2 \rangle & \mbox{for } \vert \vec k_1 \vert > k_F \mbox{ and }\vert \vec k_2 \vert > k_F \\ 0 & \mbox{else,}\end{cases}
$$
with $\vec k_i$ denoting the momenta of the interacting nucleons and the Fermi momentum $k_F$. The Bethe-Goldstone equation, however,
is more easily solved in the center of mass frame of the interacting nucleon, as the momentum of the center of mass is conserved and
the relative momentum can be expanded in a partial wave basis. Therefore one typically employs the so-called angle-average approximation for the Pauli operator and approximates the single-particle spectrum by a quadratic form
\begin{equation}
\varepsilon(\vec k) = \frac{\vec k^2}{2 m^*} + C^*\label{effmas}
\end{equation}
with an effective mass $m^*$ and a constant $C^*$ fitted to describe the single-particle spectrum for the states below the Fermi momentum. Using these approximations, the Bethe-Goldstone equation can be solved separately in each partial wave, which reduces the numerical effort drastically. Methods have been developed to treat the Pauli operator and the single-particle spectrum without these
approximations\cite{schiller,nagata} and it has been shown that the angle-average in the Pauli operator is a reasonable approximation, while the
parameterization of the single-particle spectrum according to eq.(\ref{effmas}) can lead to considerable differences, as we will also discuss below.

The problem of a precise treatment of the two-particle propagator is even more pronounced in calculation of finite nuclei, as the single-particle 
states are only defined after the corresponding Hartree-Fock equations have been solved.
Since we are using the  simple parameterization of eq.(\ref{eq:h20}) for the single-particle spectrum, we avoid a precise treatment for the 
Pauli operator ( see e.g.\cite{eigenearb}) and use the so-called angle-average for finite nuclei\cite{sauer1,sauer2} 
for a basis of oscillator states, which is appropriate for the nucleus under consideration. 

The single-particle states $\vert i,lj\rangle$ states are expanded in the very same oscillator basis $\vert n,lj\rangle_{HO}$
\begin{equation}
\vert i,lj\rangle = \sum_n c_{n,i}^{lj}\vert n,lj\rangle_{HO}\,,\label{eq:hoexpan}
\end{equation}
assuming spherical symmetry of the states for the closed-shell nuclei considered. The quantum numbers $l$ and $j$ refer to the orbital 
and total angular momenta of the single-particle states and the projection quantum numbers for the angular momentum
are dropped. An attempt is made to optimize the oscillator basis in the sense that the oscillator parameter is chosen such that the corresponding expansion
coefficients $c_{n,i}^{lj}$ are close to 1 for all occupied single-particle states. In this way it is typically sufficient to restrict the expansion 
in eq.(\ref{eq:hoexpan}) to radial quantum numbers $0\leq n\leq 4$.

The expansion coefficients $c_{n,i}^{lj}$ can then be determined by solving the BHF equations, which are given in the oscillator representation by
\begin{equation}
 \sum_n\left\lbrace \langle n',lj\vert \hat t + U^{BHF}\vert   n,lj\rangle_{HO} +  \right\rbrace c_{ni}^{lj} = \varepsilon_i  c_{n'i}^{lj}\,,\label{eq:hfequa}
\end{equation}
with $ \langle n',lj\vert \hat t \vert   n,lj\rangle_{HO}$ the matrix elements for the kinetic energy in the basis of oscillator states and the corresponding matrix elements for the BHF single-particle potential, which can be calculated as
\begin{widetext}
\begin{equation}
 \langle n',lj\vert  U^{BHF}\vert   n,lj\rangle_{HO} = 
 \sum_{kl'j'mm'J}  \frac{2J+1}{2j+1} \langle n',lj, m,'l'j'\vert G \vert   n,lj,m,l'j'\rangle_{HO}^J  \rho_k^{l'j'}c_{m'k}^{l'j'} c_{mk}^{l'j'}  \,,\label{eq:hfequa1}
\end{equation}
\end{widetext}
with the anti-symmetrized matrix-elements of the $G$-matrix $ \langle n',lj, m,'l'j'\vert G\vert   n,lj,m,l'j'\rangle_{HO}^J$ in the basis of two-nucleon oscillator states coupled to total angular momentum $J$.
The non-linear equations (\ref{eq:hfequa}) and (\ref{eq:hfequa1})  are solved in an iterative way to obtain self-consistent solutions for the expansion coefficients $c_{ni}^{lj}$ as well as single-particle energies $\varepsilon_i$ of the BHF single-particle states.

The effects of the 3N interaction are taken into account using a density-dependent 2N interaction as it is indicated in eq.(\ref{eq:v3dens}) Note, however, that the weighting coefficients of this density-dependent 2N interaction have been adjusted to obtain the correct expressions for the total energy and single-particle energies as presented in eqs.(\ref{eq:3n1}) and (\ref{eq:3n2}), respectively. 

\subsection{Rearrangement Terms}
The BHF approximation, which has briefly been sketched in the preceding subsection, corresponds to a Hartree-Fock calculation, replacing the two-particle interaction by the corresponding $G$-matrix. The $G$-matrix, however, must be understood as an effective interaction, due to its dependence on the starting energy $\omega$ and the Pauli operator, depends on the density operator of the system considered. Therefore the BHF definition of the single-particle energy does not obey the Landau definition of the quasiparticle energy in (\ref{eq:landaud}). In fact, applying the Landau prescription to
the energy functional (\ref{eq:tote}) one obtains the BHF terms of (\ref{eq:bhfeps}) plus two additional terms, the starting energy rearrangement term $\Delta U_i^\omega$ and the Pauli rearrangement term $\Delta U_i^Q$, which are due to the dependence of $G$ on starting energy $\omega$ and Pauli operator $Q$.

The starting energy rearrangement term can be written 
\begin{eqnarray}
\Delta U_i^\omega & = & \sum_{j,k}\rho_j\rho_k \langle j,k\vert \frac{\partial G}{\partial \omega}\vert j,k\rangle \frac{\partial\varepsilon_j}{\partial \rho_i} \nonumber\\
& = & \sum_{j,k}\rho_j\rho_k \langle j,k\vert \frac{\partial G}{\partial \omega}\vert j,k\rangle \langle i,j\vert G\vert i,j\rangle\,. \label{eq:startre1}
\end{eqnarray}
The second line of this equation is obtained by substituting $\varepsilon_j$ in the first line by the corresponding BHF definition of the single-particle energy. Note that adding $\Delta U_i^\omega$ to the BHF definition of the single-particle energy leads to
\begin{eqnarray}
\varepsilon_i^{RBHF} & = & \varepsilon_i + \Delta U_i^\omega \label{eq:rbhf1}\\
& = &  \langle i \vert \hat t \vert i \rangle + \sum_{j} \langle ij \vert G(\omega=\varepsilon_i + \varepsilon_j)\vert ij \rangle P_j\,,\nonumber\\
\end{eqnarray}
which means that we have replaced the single-particle density $\rho_j$ in eq.(\ref{eq:bhfeps}) by
\begin{equation}
P_j = \rho_j \left[ 1 + \sum_k \rho_k \langle j,k\vert \frac{\partial G}{\partial \omega}\vert j,k\rangle \right]\,.
\label{eq:occupation}
\end{equation}
This expression for $P_j$ typically yields values of the order of $0.8\dots 0.9$ and is often interpreted as a partial occupation of states $j$ below the Fermi energy. The approximation (\ref{eq:rbhf1}) represents the leading terms of the so-called Renormalized BHF approach (RBHF) \cite{negele,eigenearb}. Therefore we will use this name also in the following.
Note that the Koltun sum-rule of (\ref{eq:koltun}) cannot be used any longer to evaluate the total energy (\ref{eq:tote}) using the RBHF definition of the single-particle energy.

The Pauli rearrangement term can be written as
\begin{equation}
\Delta U_i^Q =- \sum_{j,k,l}\rho_j\rho_k \left\vert\langle j,k\vert G \vert i,l\rangle \right\vert^2\frac{1-\rho_l}{\varepsilon_j+\varepsilon_k-\varepsilon_i-\varepsilon_l}\,,\label{eq:paulir}
\end{equation}
and corresponds to the term of second order in $G$ in the hole-line expansion of the self-energy. Calculations including Pauli- and starting energy rearrangement terms will be denoted as density-dependent Hartree-Fock calculations (DHF) and employ single-particle energies of the form
\begin{equation}
\varepsilon_i^{DHF}  =  \varepsilon_i + \Delta U_i^\omega + \Delta U_i^Q\label{eq:dhf1}
\end{equation}

\section{Results}

\subsection{Nuclear Matter}

All the calculations presented here have been performed using the proton-neutron part of the charge-dependent Bonn interaction (CD Bonn), which has been defined and adjusted to the 2-nucleon data by Machleidt, Sammarruca and Song\cite{cdbonn}. 

\begin{figure*}[tb]
\begin{center}
\hbox{
\includegraphics[width=0.8\textwidth]{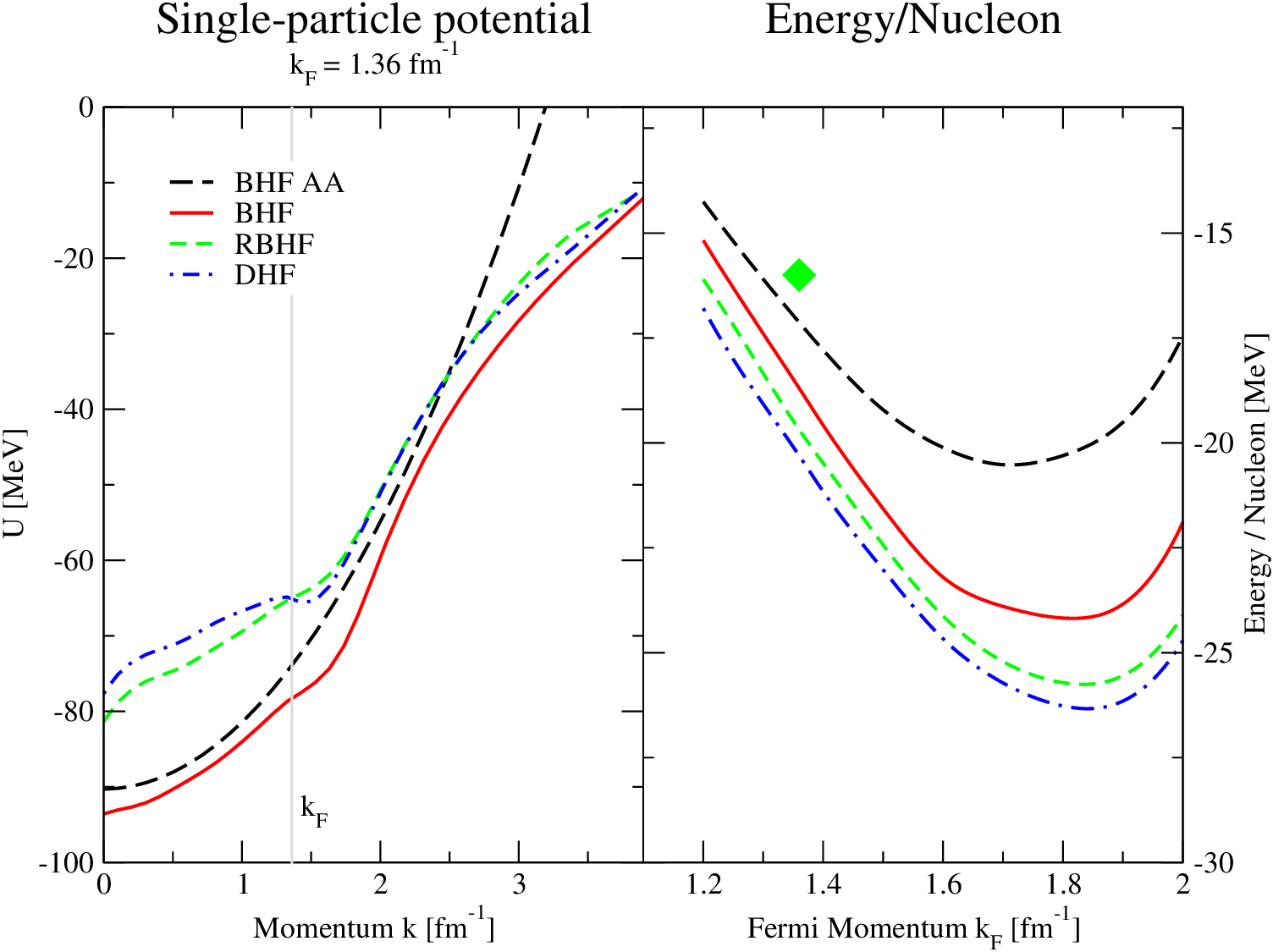}}
\caption{(Color online)  Results for symmetric nuclear matter calculations using the pn interaction of the CD Bonn potential. The left panel presents results for the single-particle potential $U(k)$ assuming a Fermi momentum $k_F$ of 1.36 fm$^{-1}$, which corresponds to the empirical saturation densities.  Results are displayed for the BHF approximation using the angle-average in the Bethe-Goldstone equation (BHF AA) and BHF, RBHF (see eq.(\protect{\ref{eq:rbhf1}})) and DHF (see eq.(\protect{\ref{eq:dhf1}})) calculations solving the Bethe-Goldstone equation without this approximation. The right panel shows the corresponding results for the energy per nucleon calculated at various Fermi momenta. }
\label{fig:nucmat1}
\end{center}
\end{figure*}

Results of conventional BHF calculations for symmetric nuclear matter, using the angle-average approximation for the Pauli operator and the parameterization of the single-particle potential using the quadratic form of eq.(\ref{effmas}) are presented by the dashed line, labeled BHF AA, in Fig.~\ref{fig:nucmat1}. The single-particle potential, which is displayed in the left panel of this figure for a Fermi momentum $k_F$ of 1.36 fm$^{-1}$, which corresponds to the empirical saturation density, reflects the quadratic parameterization, which is adjusted to reproduce the BHF single-particle potential $U(k)$ for momenta $k$ below the Fermi momentum and extended to momenta above $k_F$.  

The calculated binding energy per nucleon of such BHF AA calculations, shown in the right panel of Fig.~\ref{fig:nucmat1}, yield a minimum, representing the prediction for the saturation point, at about twice the empirical saturation density and an energy of around -20 MeV, which is much more attractive than the empirical value of -16 MeV.

The consistent treatment of the two-particle propagator in the Bethe-Goldstone eq.(\ref{eq:betheg}), avoiding the angle-average of the Pauli operator and using a consistent single-particle spectrum, leads to quite different results as can be observed from a comparison of the BHF results, presented by the red solid curves in Fig.~\ref{fig:nucmat1} and the BHF AA results. As it has been discussed by Schiller et al.\cite{schiller}, these differences can mainly be attributed to the definition of the single-particle potential. As can be seen from the left panel of Fig.~\ref{fig:nucmat1}, the single-particle energies used to define the propagator of the Bethe-Goldstone equation are quite similar for momenta below $k_F$. For the particle states with momenta above $k_F$, however the calculated BHF energies are more attractive than described by the quadratic parameterization of the BHF AA approach.

The corresponding differences in the two-particle propagator lead to matrix elements of $G$, which are in general more attractive in the BHF than in the BHF AA approach, which leads to more binding energy in the former as compared to the latter calculation. This can be seen from the energy as a function of density curves, presented in the right panel of Fig.~\ref{fig:nucmat1}. The BHF calculations yield a saturation point with even larger binding energy (-24 MeV) than the BHF AA approach at a larger saturation density.

Fig.~\ref{fig:nucmat1} also provides the relevant information about the effects of the rearrangement terms in the definition of the single-particle potential in eqs.(\ref{eq:rbhf1}) and (\ref{eq:dhf1}). The dominant contribution arises from the starting energy rearrangement term, $ \Delta U_i^\omega$, which is taken into account using the RBHF approximation. As to be expected from the representation of the RBHF energies in terms of the partial occupation probabilities defined in (\ref{eq:occupation}) these are less bound than the corresponding BHF single-particle energies. This effect is more pronounced for the states with momenta below $k_F$ than for the particle states above the Fermi momentum. This enhances the calculated binding energy even more, leading to a saturation point with an energy per nucleon of less than -25 MeV at a Fermi momentum of $k_F$ of 1.8 fm$^{-1}$, which corresponds to a saturation density 2.3 times the empirical value.

The Pauli rearrangement term, which is also included in the DHF approximation, has a small effect. Its main effect in the single-particle potential is concentrated at momenta around the Fermi momentum, where it leads to a reduction of the momentum dependence of $U(k)$. This corresponds to the enhancement of the effective mass $m^*$ to the bare mass $m$, as it has already been discussed e.g. by Mahaux and Sartor\cite{xxa} and in \cite{xxb}. The weak contribution of the Pauli rearrangement effect is also reflected in the small difference of the energies calculated in the DHF as compared to RBHF approximation.

\begin{figure*}[tb]
\begin{center}
\hbox{
\includegraphics[width=0.8\textwidth]{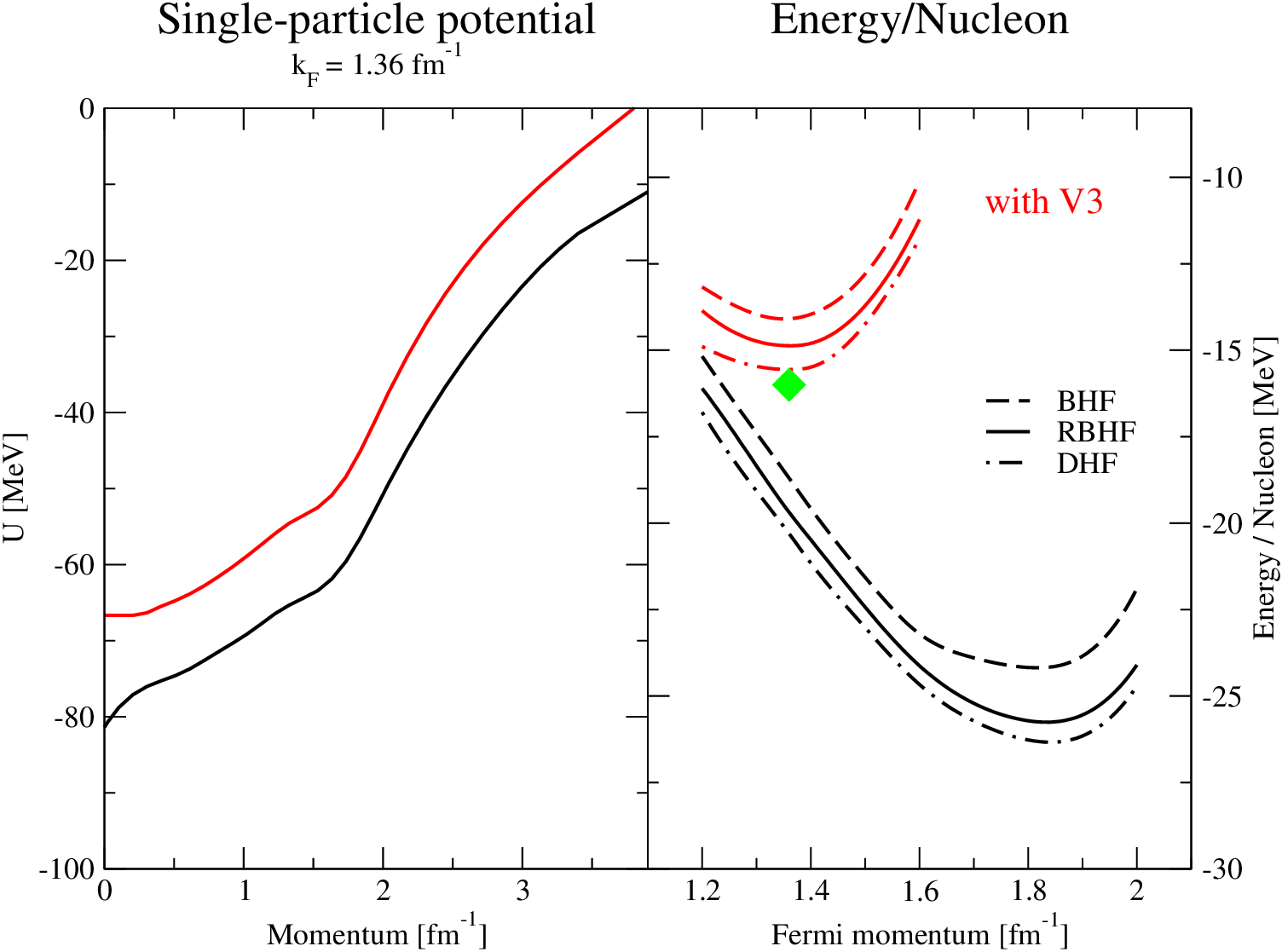}}
\caption{(Color online)  Results for symmetric nuclear matter calculations using  the CD Bonn potential with (red curves) and without inclusion (black curves) of the 3N potential. The left panel presents results for the single-particle potential $U(k)$ assuming a Fermi momentum $k_F$ of 1.36 fm$^{-1}$ RBHF approximation. The right panel shows results for the energy per nucleon calculated at various Fermi momenta for the  of BHF, RBHF and DHF approximations. }
\label{fig:nucmat2}
\end{center}
\end{figure*}

All the calculations discussed so far have been performed assuming just a realistic two-nucleon interaction, only, here the CD Bonn interaction, and we find, that the results of the calculated saturation points are part of the so-called Coester band\cite{polls,coester}. This is true for the BHF approach and the rearrangement terms just provide a shift along this Coester band. As it is one of the main goals of this investigation, to simulate the relativistic effects in terms of a  3N potential, we considered the 3N part of the Urbana interaction model\cite{urbana3,baldof} fix the parameter $A$ in eq.(\ref{eq:3-v}) to be equal zero and adjust the parameter $U$, the strength parameter for the term to simulate the change of the Dirac spinors in the medium, to reproduce the empirical saturation point. We did not aim at a high-precision fit, but just tried to obtain results close to the experimental data.

Results of such calculations are displayed in Fig.~\ref{fig:nucmat2}. It is worth noting that we can obtain a good description of the empirical saturation point by adjusting only one parameter, whereas most of the other attempts employ consider a 3 nucleon interaction with 2 or more parameters for a corresponding fit.

The 3N Term $V_{ijk}^R$ (see eq.(\ref{eq:3-v})), which we consider in our studies, is of shorter range than the corresponding 2$\pi$ exchange term ,$V_{ijk}^{2\pi}$. This may be the reason that the 3N term essentially provides a repulsive shift in the single-particle potential $U(k)$ with almost no momentum dependence (see the example in the left panel of Fig.~\ref{fig:nucmat2}). So we do not obtain the strong momentum dependence which has been observed by Zuo et al.\cite{zu074}. 

\subsection{Finite Nuclei}

\begin{table*}[tb]
\begin{tabular}{c|ccc|ccc|ccc|c}
$^{16}$O& \multicolumn{3}{c|}{NN only} & \multicolumn{3}{c|}{with 3N}&  \multicolumn{3}{c|}{with 3N GD}& Exp.\\
& \multicolumn{1}{c}{BHF} & \multicolumn{2}{c|}{RBHF}&  \multicolumn{1}{c}{BHF} & \multicolumn{2}{c|}{RBHF}&  \multicolumn{1}{c}{BHF} & \multicolumn{2}{c|}{RBHF}&\\
\hline
& $\varepsilon$ [MeV] & $\varepsilon$ [MeV] & $P$  & $\varepsilon$  [MeV] & $\varepsilon$  [MeV]& $P$ & $\varepsilon$  [MeV] & $\varepsilon$  [MeV]& $P$& $\varepsilon$\\
&{Protons}&&&&&&&& \\
$s_{1/2}$& -58.19 &-48.69&0.892 &-44.76  &-36.88  &0.917 & -41.59 & -32.89 & 0.903&-44 $\pm$ 7\\
$p_{3/2}$& -27.05 &-20.93&0.897 &-20.22 &-14.82 &0.840& -17.64 & -12.11 & 0.797 & -18.45\\
$p_{1/2}$& -20.02 &-16.25&0.871 &-16.50 &-12.27 &0.824  & -14.20 & -9.80 & 0.792& -12.12 \\
&{Neutrons}&&&&&&&&\\
$s_{1/2}$& -62.22 &-52.07&0.892 & -48.36&-39.67 & 0.918& -45.12 & -35.58 & 0.907& -47\\
$p_{3/2}$& -30.98 &-24.19&0.901 & -23.71&-17.60 &0.846 & -21.02 & -14.69 & 0.802 & -21.84\\
$p_{1/2}$& -23.83 &-19.44&0.875 & -20.00&-15.05 & 0.829 & -17.61 & -12.36 & 0.795& -15.66\\
\hline
E/A [MeV]
& -6.08 &  -6.57 & &-4.61 & -5.22&  & -3.93 & -4.81 &&-7.98 \\ 
$R_c$ [fm] & 2.35 & 2.45 && 2.59 &2.66 &  & 2.64 & 2.72 &&2.74 \\
\end{tabular}
\caption{Results for $^{16}$O using  BHF and RBHF approximation without (NN only), with inclusion of the 3N interaction (with 3N) and with 3N interaction in the global density approximation (with 3N GD). Values of single-particle energies ($\varepsilon$) occupation probabilities ($P$, see eq.\protect{\ref{eq:occupation}}) are listed for the occupied states as well as the energy per nucleon (E/A) and the radius of the charge distribution  ($R_c$). The Pauli operator in the Bethe Goldstone equation has been defined in terms of oscillator function using an oscillator parameter $b=1.767$ fm and a $C=5$  MeV has been used to define the single-particle energies in eq.(\protect{\ref{eq:h20}}).\label{VergleichO1}}
\end{table*}

The focus of the present investigation is to see if a BHF calculation with  a parameterization of the Dirac effects in terms of a 3N interaction can provide a good description for the saturation point of nuclear matter as well as the bulk properties of finite nuclei. For that purpose we performed BHF and RBHF calculations for the closed shell nuclei $^{16}$O and $^{40}$Ca using the same 2N and 3N interactions as just described for nuclear matter.

\begin{figure*}[tb]
\begin{center}
\hbox{
\includegraphics[width=0.8\textwidth]{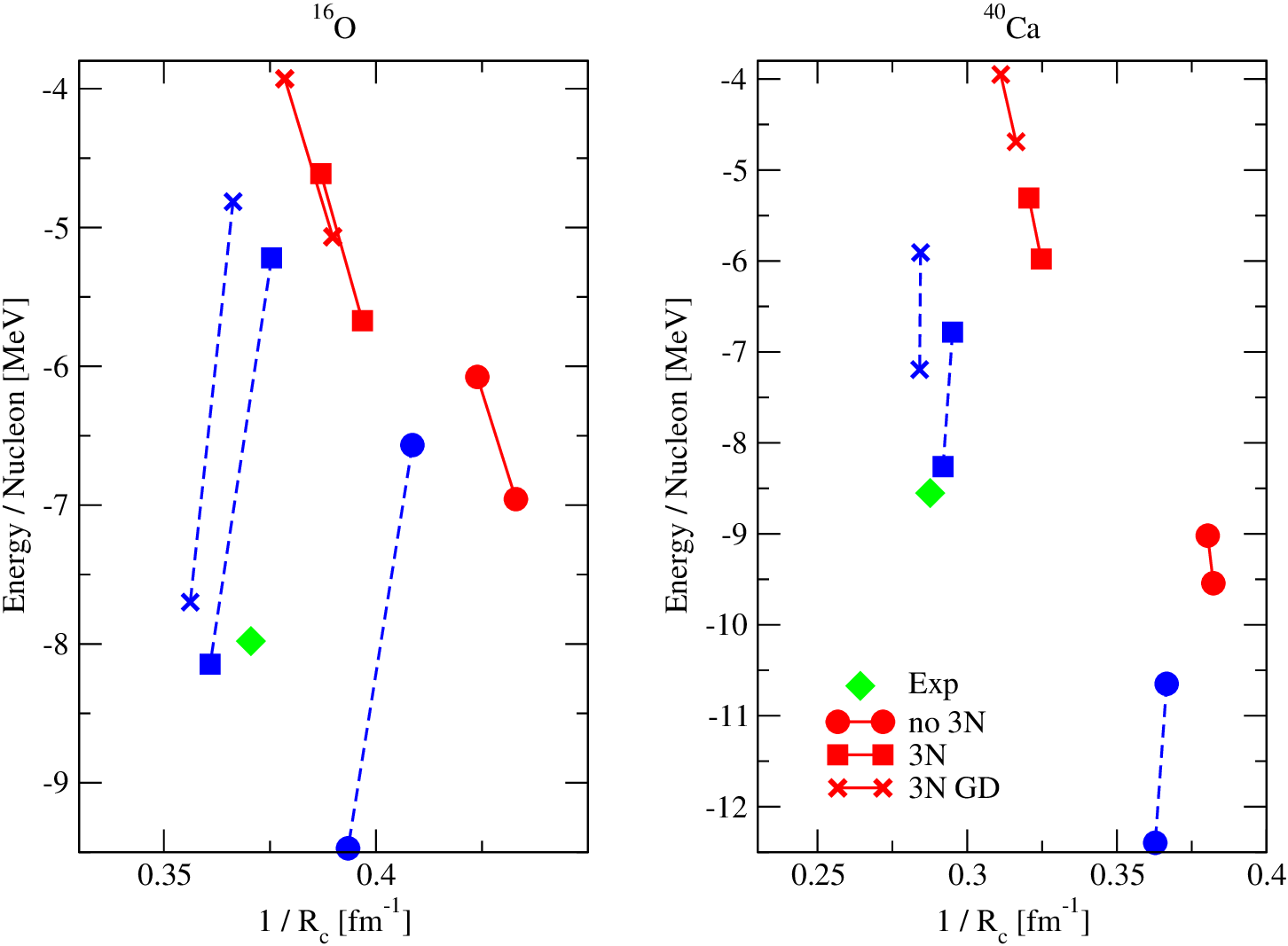}}
\caption{(Color online)  Results for the energy per nucleon and the radius of the charge distribution ($R_c$) for $^{16}$O (left panel) and $^{40}$Ca are presented in a energy versus $1/R_c$ plot to enable the comparison with the corresponding figures for nuclear matter (Fig. \protect{\ref{fig:nucmat1}} and Fig. \protect{\ref{fig:nucmat2}}).  Results referring to BHF calculations with different values of $C$ in eq.(\protect{\ref{eq:h20}}) defining the particle spectrum in the Bethe-Goldstone equation are connected by a red solid line, those of RBHF calculations are connected by a blue dashed line. Results of calculations with just the 2N interaction, with inclusion of the 3N term and including the 3N term via a global density approximation are visualized by a dot, a box and a cross, respectively. The experimental result is given in terms of a green diamond symbol. }
\label{fig:oca1}
\end{center}
\end{figure*}

Results of BHF calculations of $^{16}$O using just the CD-Bonn potential are presented in table \ref{VergleichO1}. One finds that the calculated energy per nucleon (-6.08 MeV) is less attractive as compared to the experimental value (-7.98 MeV) and the calculated radius for the charge distribution, $R_c$, is much lower (2.35 fm) than the empirical value of 2.74 fm. In order to visualize this result for the ``saturation point'' for  ${16}$O in a way, which corresponds to the plot for nuclear matter as given e.g. in the right panel of Fig.\ref{fig:nucmat2}, we indicated this result in the energy versus the inverse of the radius of the charge distribution by a red dot in Fig. \ref{fig:oca1}. In fact, it is the upper of the two red dots, connected by a solid line in the left panel  of this figure.

 Compared to the empirical data, represented by a green diamond, we observe a situation, which is quite different for  $^{16}$O than for nuclear matter (see Fig. \ref{fig:nucmat2} ). In both cases the BHF calculations yield value for $k_F$ or $1/R_c$, which are too large as compared to experiment, which implies that the average density calculated for the nuclear systems is too large. With respect to the energy per nucleon, however, the BHF calculations provide too much energy in nuclear matter and too little for the finite nucleus. Therefore in order to improve the comparison with experiment, the inclusion of the same 3N force must provide attraction in finite nuclei and repulsion in infinite matter and reduce the calculated saturation density in both cases.

The inclusion of rearrangement terms going from the  BHF to the RBHF approach yields occupation probabilities $P_i$ of the order of 0.8 to 0.9 as shown in  table \ref{VergleichO1}. From   eq.(\ref{eq:rbhf1}) it is obvious that this leads to less attractive single-particle energies in RBHF as compared to the BHF approach. The smaller attraction of the single-particle potential is reflected in a larger radius of the charge distribution. On the other hand the less attractive single-particle energies yield less attractive starting energies $\omega$  in the Bethe-Goldstone equation, which leads to more attractive matrix elements of the $G$-matrix and results in a more attractive energy per nucleon. This means that the inclusion of rearrangement terms shifts both, the energy per nucleon and the radius of the charge distribution closer to the experiment. As one can see from table  \ref{VergleichO1} and  Fig. \ref{fig:oca1} this effect is too small to provide a satisfying agreement.

The study of nuclear matter discussed above already showed that the results of BHF kind of calculations are rather sensitive to a consistent treatment of the two-particle propagator in the Bethe-Goldstone equation. The treatment of the particle-state spectrum in particular requires special attention. The same is of course also to be expected for finite nuclei. As discussed in the previous section, we approximate the Pauli operator $Q$ in the Bethe-Goldstone eq.(\ref {eq:betheg}) by a corresponding operator assuming oscillator states with an oscillator parameter $b$ of 1.76 fm  and $b$ = 2 fm in the case of $^{16}$O and $^{40}$Ca, respectively. The spectrum of the particle states, denoted by $H_0$ in  eq.(\ref {eq:betheg}) is represented by the simple parameterization of (\ref{eq:h20}), trying to adjust the constant $C$ in such a way that the resulting spectrum roughly matches the calculated single-particle energies. All the results displayed in  table \ref{VergleichO1} have been evaluated with $C =$ 5 MeV. A larger value of $C$ leads to more attraction and so, the lines displayed in   Fig. \ref{fig:oca1} indicate the range of results changing $C$ from 5 to 15 MeV, with the less attractive energy representing the $C =$ 5 MeV result. In this figure the results of BHF calculations are represented by symbols connected by a solid red line, while the symbols connected by a dashed blue line refer to the results of RBHF calculations.
The remarkable sensitivity of the calculation on this changes in the spectrum of the particle states calls for more sophisticated investigations on this issue, to obtain unambiguous results.

The effects of including the 3N force can be seen by comparing the results visualized in  in Fig.~ \ref{fig:oca1} in terms of the square boxes with the corresponding results displayed by circles. As to be expected, the inclusion of the 3N yields a repulsive effect and leads to a reduction of the total energy accompanied by an increase of the nuclear radius. Comparing this effect with the corresponding repulsive effect one can obtain with a lowering of the parameter $C$ just discussed, one finds that the 3N force yields a larger increase in the radius, if the energy is changed by a similar amount. This may reflect the fact that a proper treatment of the 3N force as compared to a modification of the 2N interaction has a larger effect on the single-particle potential (see the discussion in the introduction connected to eqs.(\ref {eq:3n2}) and  (\ref {eq:v3dens}).
Therefore a repulsive 3N interaction will be more efficient in changing the single-particle potential and the resulting radius of the particle distribution than a similar change in the 2N interaction.

 Fig. \ref{fig:oca1} also shows results displayed in terms of crosses and denoted as 3N GD (see also table \ref{VergleichO1}) . As discussed above, this approximation scheme describes an attempt to evaluate the effects of the 3N force in nuclear matter at various densities and then transfer the result to the calculation of finite nuclei by choosing the density of nuclear matter to be identical to the average particle density calculated  for the nucleus considered. We find that this global density (GD) approximation yields effects which are similar with a tendency to overestimate the corresponding effects of a direct treatment in the finite nucleus  by 20 to 30 percent.

\begin{table*}[tb]
\begin{tabular}{c|ccc|ccc|ccc}
& \multicolumn{3}{c|}{NN only} & \multicolumn{3}{c|}{with 3N}&  \multicolumn{3}{c}{with 3N GD}\\
& \multicolumn{1}{c}{BHF} & \multicolumn{2}{c|}{RBHF}&  \multicolumn{1}{c}{BHF} & \multicolumn{2}{c|}{RBHF}&  \multicolumn{1}{c}{BHF} & \multicolumn{2}{c}{RBHF}\\
\hline
& $\varepsilon$ [MeV] & $\varepsilon$ [MeV] & $P$  & $\varepsilon$  [MeV] & $\varepsilon$  [MeV]& $P$ & $\varepsilon$  [MeV] & $\varepsilon$  [MeV]& $P$\\
&{Protons}&&&&&&& \\
$0s_{1/2}$& -94.27 &-76.04&0.850 &-53.18&-36.65&0.855& -45.35&-31.34&0.790 \\
$p_{3/2}$& -63.36 &-48.11&0.869 &-34.63&-21.93&9.745& -27.58&-17.92&0.692 \\
$p_{1/2}$& -53.82 &-41.61&0.863 &-30.53&-19.82&0.723 & -24.10&-16.16&0.676 \\
$d_{5/2}$&-33.20&-21.81&0.816&-16.61&-8.47&0.718 &-11.36&-5.51&0.724 \\
$1s_{1/2}$&-26.61&-18.06&0.739&-12.88&-7.15&0.726 &.9.33&-5.04&0.743 \\
$d_{3/2}$&-18.22&-12.20&0.736&-10.66&-5.29&0.713 &-6.33&-2.85&0.731 \\
&{Neutrons}&& &&&&&\\
$s_{1/2}$& -104.3 &-83.71&0.854 & -61.35&-42.02&0.868 & -53.30&-36.39&0.810 \\
$p_{3/2}$& -72.74 &-55.31&0.869 & -42.44&-27.02&0.760  & -35.02&-22.71&0.704 \\
$p_{1/2}$& -63.12 &-48.74&0.865 &-38.23&-24.84&0.737 & -31.41&-20.89&0.687 \\
$d_{5/2}$&-42.16&-28.58&0.830&-23.99&-13.36&0.720 &-18.38&-10.13&0.722 \\
$1s_{1/2}$&-34.93&-24.43&0.752&-20.13&-12.00&0.723 &-16.26&-9.63&0.736 \\
$d_{3/2}$&-26.88&-18.72&0.749&-17.84&-10.09&0.712 &-13.19&-7.39&0.725 \\
\hline
E/A [MeV]
& -9.02 &  -10.65 & &-5.31 & -6.78&  & -3.95 & -5.91 & \\ 
$R_c$ [fm] & 2.62 & 2.72 && 3.11 &3.38 &  &3.21 & 3.52 & \\
\end{tabular}
\caption{Results for $^{40}$Ca using  BHF and RBHF approximation. The Pauli operator in the Bethe Goldstone equation has been defined in terms of oscillator function using an oscillator parameter $b=2.0$ fm and a $C=10$  MeV has been used to define the single-particle energies in eq.(\protect{\ref{eq:h20}}). Further details see caption of table \protect{\ref{VergleichO1}}\label{VergleichCA1}}
\end{table*}

Similar calculations have also been done for the nucleus $^{40}$Ca. Results on the ``saturation properties'' of this nucleus are displayed in the right panel of  Fig. \ref{fig:oca1} considering  the values $C$ = 10 and $C$ = 15 MeV for the parameterization of the particle state spectrum. More explicit results on the single-particle energies are shown in table \ref{VergleichCA1} assuming $C$ = 10 MeV. The main features of these results for  $^{40}$Ca are very similar to those discussed for the example $^{16}$O and therefore confirm these findings.

It is worth noting that, assuming $C$ = 15 MeV, and including the 3N force RBHF calculations of both nuclei yield results for energy and radius of the charge distribution, which are in a good agreement with the experimental data (see  Fig. \ref{fig:oca1} and table \ref{Vergleichexp}). We do not intent to celebrate this as a success of the Dirac BHF approach, or BHF approach simulating the Dirac effects of DBHF in terms of a 3N force. We will keep in mind that the strength of the 3N force, which was motivated to simulate the Dirac effect, has been adjusted to reproduce the saturation point of nuclear matter. With the 3N force adjusted we made a reasonable but not uniquely justified choice for the description of the particle-state spectrum to end up with a good description of the bulk properties of finite nuclei as well.

Nevertheless, this result shows that using an appropriate 3N force to simulate the effects of Dirac spinors modified in the nuclear medium within the framework of non-relativistic BHF calculations, one may be able to describe the bulk properties of nuclear matter and finite nuclei based on a realistic NN interaction.

The description of bulk properties (energy, radius of particle distribution, density) is an important but only one of features of nuclear structure, which one hopes to describe within the relativistic DBHF. Other important aspects one hopes to describe within a relativistic description of nuclear systems are the energy dependence of the optical potential\cite{ruirui} and the spin-orbit splitting of the single-particle energies\cite{Zamick}, which is enhanced due to the enhancement of the small component of Dirac spinors in the nuclear medium.

Is this enhancement of the spin-orbit splitting, which is important to describe the strength of the spin-orbit term observed in the experiment, also simulated by the simple 3N force of eq.(\ref{eq:3-v})? Inspecting e.g. the single-particle energies of the $p_{3/2}$ and $p_{1/2}$ states listed in table \ref{VergleichO1} we do not find an enhancement of the spin-orbit splitting with the 3N force included. In fact, the differences between these single-particle energies are always smaller with inclusion of the 3N force. This is related to the fact that the 3N force yields larger values for the radii which reduces the spacing between the single-particle states. But even, if one accounts for this size effect, the 3N force does not provide an enhancement of the spin-orbit splitting. This could be achieved by introducing an appropriate spin structure in the 3N force, a feature which presumably would lead to more parameters and spoil the simplicity of the present approach.

\begin{table}[tb]
\begin{tabular}{cc|cc|cc|c}
&& \multicolumn{2}{c|}{NN only} & \multicolumn{2}{c|}{with 3N}&  \\
&& BHF & RBHF & BHF & RBHF & Exp \\
\hline
$^{16}$O & E/A [MeV] & -6.96 & -9.47 & -5.67 & -8.15 & -7.98\\
& $R_c$ [fm] & 2.31 & 2.54 & 2.52 & 2.77 & 2.70 \\
\hline
 & E/A [MeV] & -9.54 & -12.40 & -5.98 & -8.26 & -8.55 \\
& $R_c$ [fm] & 2.62 & 2.76 & 3.08 & 3.43 & 3.48 \\
\end{tabular}

\caption{Results for the energy per nucleon $E/A$ and the radius of the charge distribution $R_c$ for $^{16}$O and $^{40}$Ca calculated in BHF and RBHF approximation are compared to the experimental data \protect{\cite{exp1,exp2}}. In contrast to the calculations leading to the results in table \protect{\ref{VergleichO1}} and table \protect{\ref{VergleichCA1}} a shift
$C=15$  MeV has been used to define the single-particle energies in eq.(\protect{\ref{eq:h20}}).\label{Vergleichexp}}
\end{table}

\section{Conclusion}
An attempt has been made to simulate the relativistic features of the Dirac-Brueckner-Hartree-Fock (DBHF) approach by adding effects of a simple 3 nucleon (3N) force to non-relativistic many-body calculations based on the BHF approach.
One parameter, the strength of the 3N force, is adjusted to reproduce the empirical saturation point of infinite nuclear matter and than used without further modifications for the description of finite nuclei. Special attention is paid to the energy spectrum of particle states, which is used in the propagator of the Bethe-Goldstone equation, and the importance of rearrangement terms due to the energy dependence of the effective interaction. For the studies of finite nuclei in particular more effort is needed to optimize the treatment of the propagator in the Bethe-Goldstone equation. Taking a reasonable choice for the particle state spectrum and including the effects of rearrangement terms the Renormalized Brueckner Hartree Fock calculations with 3N force can reproduce the empirical values for the energy and radius of charge distribution of nuclei like $^{16}$O and $^{40}$Ca. This attempt to simulate the effects of the relativistic features of DBHF, however, fails to reproduce other predictions of the Dirac phenomenology like the strength of the spin-orbit term in the single-particle field.

\section{Acknowledgments}
This work has been performed as a part of a project (Mu 705/10-1) supported by the Deutsche
Forschungsgemeinschaft, DFG.


\begin{thebibliography}{99}
\bibitem{skyrme1} T.H.R. Skyrme, Nucl. Phys. {\bf 9}, 615 (1959).
\bibitem{skyrme2} J. Sadoudi, T. Duguet, J. Meyer, and M. Bender, Phys. Rev. C {\bf 88}, 064326 (2013).
\bibitem{walecka} B. D. Serot and J. D. Walecka, Adv. Nucl. Phys. {\bf 16}, 1 (1986).
\bibitem{mach0} R. Machleidt, Adv. Nucl. Phys. {\bf 19}, 189 (1989).
%
\bibitem{V18} R.B. Wiringa, V.G.J. Stoke, and R. Schiavilla, Phys. Rev. C {\bf 51}, 38 (1995).
%
\bibitem{cdbonn} R. Machleidt, F. Sammarruca, and Y. Song, Phys. Rev. C {\bf 53}, R 1483 (1995).
%
\bibitem{polls} H. M\"uther and A. Polls, Prog. Part and Nucl. Phys. {\bf 45}, 243 (2000).
%
\bibitem{negele} K.T.R. Davies, R.J. McCarthy, J.W. Negele, and P.U. Sauer, Phys. Rev. C {\bf 10}, 2607
(1974).
%
\bibitem{eigenearb} R.K. Tripathi, A. Faessler, and H. M\"uther, Phys. Rev. C {\bf 10}, 2080
(1974).
%
\bibitem{david} K.T.R. Davies and R.J. McCarty, Phys. Rev. C {\bf 4}, 81 (1971).
%
\bibitem{zuo} W. Zuo, I. Bombaci, and U. Lombardo, Phys. Rev. C {\bf 60}, 024605 (1999).
%
\bibitem{Baymk} G. Baym and L.P. Kadanov, Phys. Rev. {\bf 124}, 287 (1961).
%
\bibitem{coester} F. Coester, S. Cohen, B.D. Day, and C.M. Vincent, Phys. Rev. C {\bf 1}, 769 (1970). 
%
\bibitem{holinde} K. Holinde Phys. Rep. {\bf 68}, 121 (1981).
%
\bibitem{wiringa} R.B. Wiringa, R.A. Smith, and T.L. Ainsworth, Phys. Rev. C {\bf 29}, 1207 (1984).
%
\bibitem{delta1} M.R. Anastasio, H. M\"uther, A. Faessler, K. Holinde, and R. Machleidt,
Phys. Rec. C {\bf 18}, 2916 (1978).
%
\bibitem{delta2} A. Faessler, H. M\"uther, K. Shimizu, and W. Wadia, Nucl. Phys. {\bf A333}, 428 (1980).
%
\bibitem{dirac1} R. Brockmann and R. Machleidt, Phys. Rev. C {\bf 42}, 1965 (1990). 
%
\bibitem{dirac2} R. Fritz, H. M\"uther, and R. Machleidt, Phys. Rev. Lett. {\bf 71}, 46 (1993).
\bibitem{brown87} G.E. Brown, W. Weise, G. Baym, and J. Speth, Comments Nucl. Part. Phys.
\textbf{17}, 39 (1987).
%
\bibitem{Bouyssy87} A. Bouyssy, J.-F. Mathiot, N. van Giai, and S. Marcos, Phys. Rev. C
\textbf{36}, 380 (1987).
\bibitem{anastasio83} M.R. Anastasio, L.S. Celenza, W.S. Pong, and C.M. Shakin, Phys. Rep.
\textbf{100}, 327 (1983).
%
\bibitem{horow87} C.J. Horowitz and B.D. Serot, Nucl. Phys. A \textbf{464}, 613 (1987).
%
\bibitem{terhaar:1987} B. ter Haar and R. Malfliet,
Phys. Rep. \textbf{149}, 207 (1987).
%
\bibitem{dejong:1998}
F. de Jong and H. Lenske,
Phys. Rev. C \textbf{58}, 890 (1998).
%
\bibitem{gross:1999}
T. Gross-Boelting, C. Fuchs, and Amand Faessler,
Nucl. Phys. \textbf{A648}, 105  (1999).
%
\bibitem{alonso:2003} D. Alonso and F. Sammarruca, Phys. Rev. C \textbf{67}, 054301 (2003).
%
\bibitem{eric} E.N.E. van Dalen and H. M\"uther, Int. J. Mod. Phys. E \textbf{19}, 2077 (2010).
%
\bibitem{giai2010} N. van Giai, B.V. Carlson, Z. Ma, and H.H. Wolter, J.Phys. G \textbf{37},
064043 (2010).
%
\bibitem{releric} E.N.E. van Dalen and H. M\"uther, Phys. Rev. C {\bf 84}, 024320 (2011).


\bibitem{fuchs95} C. Fuchs, H. Lenske, and H.H. Wolter, Phys. Rev. C \textbf{52}, 3043 (1995).

\bibitem{brock88}  H. M\"uther, R. Machleidt, and  R. Brockmann, Phys. Rev. C \textbf{42}, 1981 (1990).


%
\bibitem{urbana3} J. Carlson, V.R. Pandharipande, and R. Wiringa, Nucl. Phys. {\bf A401}, 59 (1983).

\bibitem{baldof} M. Baldo and L. Ferreira, Phys. Rev. C {\bf 59}, 682 (1999).
%
\bibitem{grange} P. Grange, A. Lejeune, M. Martzolff, and J.-F. Mathiot, Phys. Rev. C {\bf 40}, 1040 (1989).
%
\bibitem{soma} V. Soma and P. Bo\.{z}ek, Phys. Rev. C {\bf 78}, 054003 (2009).
%
\bibitem{hebschw} K. Hebeler and A. Schwenk, Phys. Rev. C {\bf 82}, 014314 (2010). 
%
\bibitem{carbone} A. Carbone, A. Rios, and A. Polls, Phys. Rev. C {\bf 88}, 044302 (2013).
%

\bibitem{koltun} D.S. Koltun, Phys. Rev. Lett {\bf 28}, 182 (1972).

\bibitem{song} H.Q. Song, M.G.Baldo, Giansiracusa, and U. Lombardo, Phys. Rev. Lett. {\bf 81}, 1584 (1998).

\bibitem{jeuken76} J.P. Jeukenne, A. Lejeunne, and C. Mahaux, Phys. Rep. {\bf 25}, 83 (1976).  

\bibitem{schiller} E.~Schiller, H.~M\"uther and P. Czerski, Phys. Rev. C {\bf 59}, 2934 (1999),
 Erratum,  Phys. Rev. C {\bf 60}, 059901 (1999).

\bibitem{nagata} K. Suzuki, R. Okamoto, M. Kohno, and S. Nagata, Nucl. Phys. A {\bf 150}, 467 (1970).

\bibitem{sauer1} P.U. Sauer, Nucl. Phys. A {\bf 665}, 92 (2000).

\bibitem{sauer2} H. M\"uther and P.U. Sauer, Computational Nuclear Physics 2, (Springer Verlag N.Y. 1993) 30.

\bibitem{xxa} C. Mahaux and R. Sartor, Adv. Nucl. Phys. {\bf 20}, 1 (1991).

\bibitem{xxb} Kh.S.A. Hassaneen and H, M\"uther, Phys. Rev. C {\bf 70}, 054308 (2004).

\bibitem{zu074} W. Zuo, U. Lombardo, H.-J. Schulze, and Z.H. Li Phys. Rev. C {\bf 74}, 014317 (2006).

\bibitem{exp1} Atomic Mass Data of National Nuclear Data Center, Brookhaven National Laboratory, www.nndc.bnl.gov

\bibitem{exp2} I. Angeli and K.P. Marinova, At. Data Nuc. Data Tables {\bf 99}, 69 (2013).

\bibitem{ruirui} Ruirui Xu, Zhongyu Ma,  E.N.E. van Dalen, and H. M\"uther, Phys. Rev. C {\bf 85}, 034613 (2012).

\bibitem{Zamick} L. Zamick, D,C. Zheng, and H. M\"uther, Phys. Rev. C {\bf 45}, 2763 (1992).

\end{thebibliography}
\end{document}